# A Novel Design for Cruiser Type Motorcycle Silencer Based on Micro-Perforated Elements

**Kabral, R.; Rämmal, H.; Auriemma, F.; Luppin, J.; Koiv, R.; Tiikoja, H.; Lavrentjev, J.**

Tallinn University of Technology, Tallinn, Estonia

## ABSTRACT

Regulations stipulating the design of motorcycle silencers are strict, especially when the unit incorporates fibrous absorbing materials. Therefore, innovative designs substituting such materials while still preserving acceptable level of characteristic sound are currently of interest.

Micro perforated elements are innovative acoustic solutions, which silencing effect is based on the dissipation of the acoustic wave energy in a pattern of sub-millimeter apertures. Similarly to fibrous materials the micro perforated materials have been proved to provide effective sound absorption in a wide frequency range. Additionally, the silencer is designed as a two-stage system that provides an optimal solution for a variety of exploitation conditions.

In this paper a novel design for a cruiser type motorcycle silencer, based on micro-perforated elements, is presented. It has been demonstrated that the micro perforated elements can successfully be used to achieve high attenuation of IC-engine noise in strictly limited circumstances. A technical description of the design and manufacturing of the prototype silencer is given and technological issues are discussed. The acoustical and aerodynamical performance of the silencer is characterized by transmission loss and pressure drop data. The influence of the two-stage system valve operation has been analyzed by studying the acoustics data and engine output characteristics.

In addition to the experimental investigations, numerical 1-D models were developed for the optimization of the silencer geometry and the results are compared in a number of operating conditions.

The studies have resulted in development of a silencer system for a small series cruiser type motorcycle. The first silencer prototypes have been tested on the motorcycle. While maintaining acceptable pressure drop characteristics, it has proven to comply with standard noise criteria without incorporating fibrous materials.

The radiated motorcycle sound, as one of the key features of successful design, has been evaluated. The sound design has been recognized as well suitable for the product.

## INTRODUCTION

A motorcycle design is a complex task integrating challenges of engineering analyzes with marketing goals while complying with a number of regulations. The exhaust system of this type of motorcycle (see Fig. 1) has to provide not only adequate engine noise cancellation but also a pleasant characteristic sound while preserving acceptable level of exhaust gas flow restriction.

The present work is focusing on the exhaust system design for exclusive small series motorcycle. Traditionally, this type of motorcycle is equipped with a relatively large displacement internal combustion engine (see Table 1) designed for a moderate rpm range.

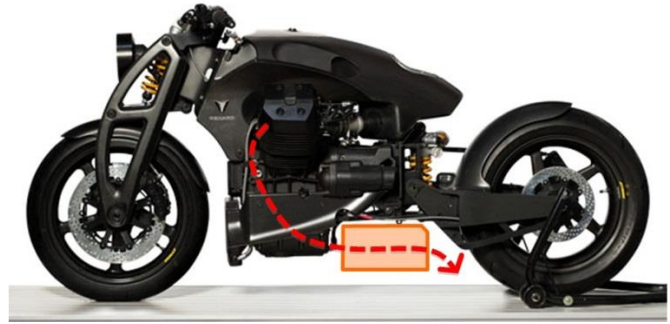

*Figure 1 – A side view of the motorcycle prototype [1]. The location of the silencer unit (orange rectangle) and the direction of exhausting gas flow (dashed red arrow) are illustrated.*

*Table 1 – Characteristic parameters of the motorcycle engine [1].*

| layout | V2, 90° |
|---|---|
| total displacement | 1326cm$^3$ |
| geom. compression ratio | 10,8 |
| bore | 102mm |
| stroke | 81,2mm |
| max. torque | 134Nm (5600rpm) |
| max. power | 123hp (7100rpm) |
| engine management | electronical (EFI, Euro4) |
| cooling system | water cooling |

In order to introduce this type of vehicle for initial registration in European Union (EU) according to Directive 2002/24/EC [2] the type approval evaluation has to be performed. Procedures for the evaluation of exhaust systems are stipulated in Directive 97/24/EC [3]. Accordingly the maximum by-pass (vehicle in motion) noise level for this type of motorcycle is limited to 80dB(A).

Typically, a motorcycle silencer incorporates fibrous materials. To avoid negative side effects, e.g. deterioration of performance due to relocation and blow out of the fibers, challenging demands have been stipulated in Directive 97/24/EC [3] on the implementation of the fibrous materials. A recent investigation [4] on the use of micro-perforated (MP) elements in flow-duct silencers has proven that the fibrous materials can be successfully substituted by the MP elements.

As a well-known issue the noise attenuation ability typically compromises the pressure drop of the silencer. As the pressure drop is one of the key parameters affecting the engine performance and fuel consumption, innovative technical solutions (e.g. micro perforations and meta-materials) providing satisfactory noise cancellation while preserving low pressure drop are of great interest.

The motorcycle exhaust system treated in this paper was designed taking into account the following technical constraints, concentrated into Table 2. In the present work a technical description of the silencer is given together with experimentally and numerically obtained characteristic parameters, relevant to successfully fulfill the technical constraints listed in Table 2.

*Table 2 – Technical constraints for the exhaust system design.*

| No. | Constraint |
|---|---|
| 1 | The geometry of the silencer is restricted by the overall design of the motorcycle components (See Fig. 1). |
| 2 | The motorcycle equipped with the silencer must satisfy the stipulated noise limits [3]. |
| 3 | The silencer design should provide an option for less flow restrictive "straight flow" configuration. |
| 4 | The fibrous acoustic materials should be avoided [3] inside the silencer. |
| 5 | The silencer should resist corrosive environments |
| 6 | The mass of the complete silencer system should be minimal. |

# SILENCER DESIGN

## OVERVIEW OF DESIGN PROCEDURES

The research and development of the silencer system including experimental testing and 1-D computer simulations were carried out in cooperation with technical acoustics laboratory at Tallinn University of Technology. A dedicated hot flow test facility presented in [4, 5 and 6] was implemented for the experimental acoustic characterization of the complete silencer and silencer components.

During the product development project the following design procedures were performed:

1. Analysis on the positioning of the silencer. Determination of the space available, the location of the mounting structures and the inlet and outlet tubes;
2. Manufacturing and acoustic characterization of micro-perforated tubular elements (40 test samples);
3. Set up of acoustic simulation models in 1-D analysis software followed by preliminary analyses;
4. Manufacturing and acoustic characterization of the first geometrically simplified silencer prototype;
5. On-vehicle testing of the silencer prototype in a variety of operating conditions (stationary: idle, 3000rpm, 6000rpm and by-pass tests [3]);
6. Development and road testing [3] of extra noise control guide valves;
7. Motorcycle testing on rolling road and analysis for the engine equipped with the 2-stage silencer prototype;
8. Acoustic 2-port testing and characterization of noise control guide valves;
9. Manufacturing of the complete silencer system for the motorcycle.

## ACOUSTICAL DESIGN

Since the dominant noise radiation typically originates from the exhaust gas pressure pulsations related to the first harmonics of the firing frequencies, the engine (see Table 1) tuned for medium crankshaft RPM range can be regarded as a low frequency source. Hence, an effective noise cancellation for such engine is technically obtained by the maximization of the acoustic wave reflections. In order to provide high reflections the cross section area ratio at the sudden exhaust system area discontinuities (e.g. in expansion chambers) should be maximized. Therefore, it is natural to utilize all the limited volume available. A CAD model of the silencer presented (Fig. 2) illustrates an effective use of the space available underneath the motorcycle (see Fig. 1).

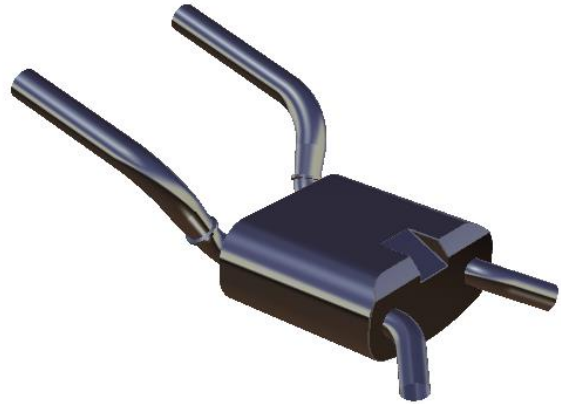

*Figure 2 – A CAD model of the geometry available for the exhaust system.*

Acoustically the silencer (see Fig. 3) has been designed to incorporate a combined noise cancellation principle, where the exhausting pressure pulsations are reflected backwards to the source by the three sequential reactive expansion chambers (a, b and c) and attenuated by the dissipative micro-perforated tubes (1 and 2). Both the engine cylinders are designed to exhaust via autonomous primary pipes and tailpipes, which are coupled inside the silencer housing through the micro-perforation. In order to satisfy the described noise limits the perforated tubes inside the silencer have been equipped with guide valves (see Fig. 4), positioned in the middle of the largest expansion chamber. The function of the guide valves is to direct the pulsating exhausted gas flow through the micro-perforated elements before terminating from the tailpipe. The dissipative acoustic effect of the micro-perforated elements has been found to be remarkably dependent on the viscous losses introduced inside the perforated apertures [6, 7 and 8] and the sound dissipation typically improves in higher flow velocity conditions. The reduction of propagating sound pressure amplitudes can be explained by the increase in acoustic losses originating from the flow induced vortex shedding in this region. Hereby, this technical solution has acoustically and aerodynamically been proved to successfully enhance the noise cancellation while preserving acceptable pressure drop (see results section).

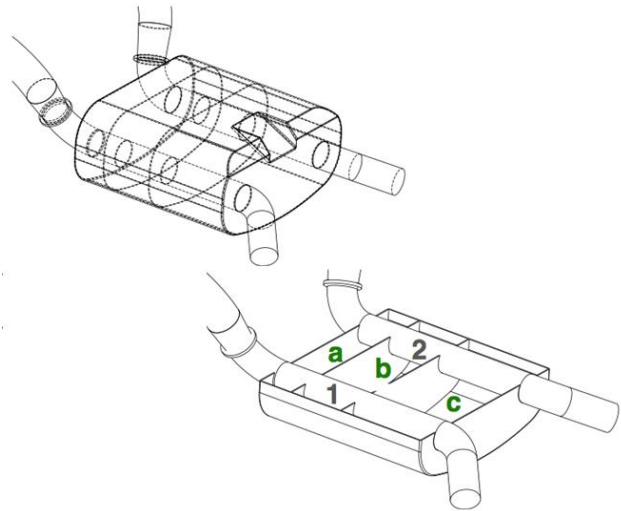

*Figure 3 – A geometrical layout of silencer unit revealing the three sequential expansion chambers: 1, 2 – micro-perforated pipes and a, b, c – expansion chambers.*

# TECHNICAL DESCRIPTION

The technical layout of the silencer incorporating three sequential expansion chambers and micro-perforated tubes is described in Figs. 2-4 and the characteristic data of the micro-perforated tubular elements is presented in Table 3.

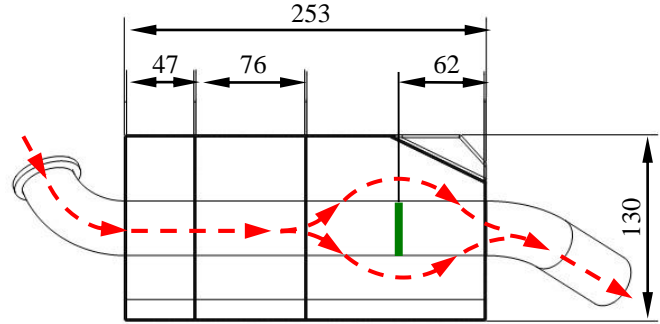

*Figure 4 – A side view of the silencer unit exhibiting removable guide valve position (green line) and illustrating the gas flow path along the ducts and through the micro-perforated walls (dashed red arrows).*

*Table 3 – Characteristic data of the micro-perforation used in the silencer.*

| | |
|---|---|
| **Photo** | 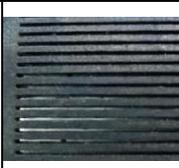 |
| **Aperture sketch** | 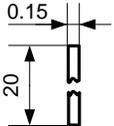 |
| **Porosity** | 10% |
| **PA ratio** | 13.4 |

# TECHNOLOGY AND MANUFACTURING

An overview of the materials and technological procedures used to manufacture the silencer system components is composed in Table 2.

Preprint of the paper: Kabral, R.; Rämmal, H.; Auriemma, F.; Luppin, J.; Koiv, R.; Tiikoja, H.; Lavrentjev, J. "A Novel Design for Cruiser Type Motorcycle Silencer Based on Micro-Perforated Elements".
SAE Technical Paper 2012-32-0109, 2012. https://doi.org/10.4271/2012-32-0109*Table 4 – Material specifications and technological procedures used for the manufacturing of the silencer components.*

| Silencer component | Quantity | Parts, material | Technological procedures |
|---|---|---|---|
| Inlet ducts | 2 | Pipe with a bend, stainless steel AISI316 (45x1.25-R100)<br><br>conical element, stainless steel AISI316 (D45-D38x1.25-L44.1)<br><br>2 flanges, V-clamp<br><br>Pipe with a bend, stainless steel AISI316 (38x1.25-R57) | Cut, machined, welded* |
| Perforated duct | 2 | Pipe, stainless steel AISI316 (38x1.25) | Laser perforated |
| Outlet ducts | 2 | Pipe, stainless steel AISI316 (38x1.25) | Cut, welded* |
| Silencer housing | 1 | Sheet metal, stainless steel AISI304 (1.5mm) | Laser cut, formed, welded* |
| Baffles | 3 | Sheet metal, stainless steel AISI304 (1.5mm) | Laser cut |
| Guide valve | 2 | Rod, stainless steel AISI316 (D40) | Machined |

*TIG welding

## EXPERIMENTS

In order to evaluate the overall performance of the silencer a variety of common engineering parameters were determined including acoustic transmission loss spectra, the vehicle in motion noise emission level and the aerodynamic pressure loss (see the results section). A symmetrical geometry of the silencer unit (see Fig. 3) is expected to offer equal TL and PD characteristics for both exhaust passages (sides). To simplify the experimental procedures, according to the assumption, the characteristic data were determined only for one side of the silencer.

Page 7 of 12

For the acoustic characterization the silencer unit was treated as an acoustic two-port [9]. The acoustic power was determined at the inlet and outlet cross-section of the silencer by using the classical two-microphone wave decomposition method [10]. The TL as the measure of the acoustic power reduction across the silencer was calculated [11]:

$$\text{TL} = 10 \log\left(\frac{W_{\text{in}}}{W_{\text{out}}}\right) =$$

$$= 10 \log\left[\frac{A_{\text{in}} \cdot \rho_{\text{out}} \cdot c_{\text{out}} \cdot (1 + M_{\text{in}})^2}{|T|^2 \cdot A_{\text{out}} \cdot \rho_{\text{in}} \cdot c_{\text{in}} \cdot (1 + M_{\text{out}})^2}\right], \quad (1)$$

where $W$ denotes the acoustic power, $A$ is the cross sectional area of the flow-duct, $\rho$ is the density of the flowing media, $c$ is the speed of sound, $M$ is a Mach number and $T$ is the transmission coefficient [11]. The subscripts denote the respective sides of the two-port. The transmission coefficient $T$ originates from the scattering matrix formulation [10] of the two-port and is defined as the ratio between incident and transmitted complex acoustic wave amplitudes. In [5] a detailed description about the determination of the transmission coefficient $T$ has been treated by the authors together with the overview of the aero-acoustic hot flow facility used for the experimental investigations in this paper.

The maximum by-pass (vehicle in motion) noise level was measured following the Directive 97/24/EC [3] and the aerodynamic PD was obtained by measuring the static pressure difference across the inlet and outlet of the silencer.

# MODELLING

The 1D analysis of the muffler has been performed by implementing commercial software packages Gamma Technologies GT-Power[TM] [12] and SIDLAB[TM] [13]. GT-Power[TM] is a widely employed tool for the analyses of internal combustion engines, and the SIDLAB[TM] is a specific tool for 1D acoustic simulations focusing on flow duct applications. In GT-Power[TM] the 1D non-linear flow equations of mass, momentum and energy conservation [14] are solved in the time domain for each component of the complete engine (air box, intake, exhaust system, muffler, etc.).

In the specific case of a silencer, the chambers are schematized by sets of quasi-3D elements whose ports are oriented along the three spatial directions. In this way the radial interactions between the perforated tubes and the chambers are accounted. Straight tubes are represented by 1D elements. Eventually, perforated tubes are modeled by a set of quasi-3D elements connected to each other along the direction of the tube and linked, through a proper number of orifices, to the quasi-3D elements constituting the chambers (see Fig. 5).

The alternative software package SIDLAB uses a 1D linear frequency domain approach, which is very fast and especially suitable for flow duct acoustic applications. The computations are based on the well-known acoustic plane wave equation [15]. In GT-Power[TM] the porosity is accounted considering a respective friction coefficient in the energy equation. Whereas in Sidlab the perforated tubes are divided into a number of discrete lumped impedance two-port elements which are separated by hard segments on both sides [16]. The two-microphone random-excitation technique proposed by Seybert and Ross [17] is used in GT-Power[TM] in order to calculate the transmission loss of the silencer.

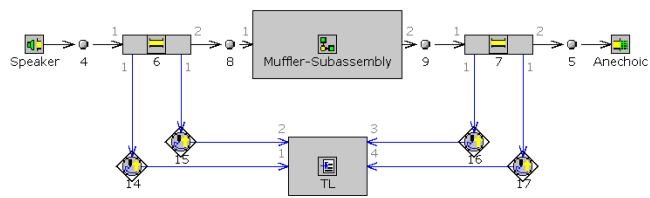

*Figure 5 – The two-microphone technique virtually implemented for the TL analysis in 1D GTPower simulations.*

# RESULTS

## EXPERIMENTAL RESULTS

Experimentally determined results for the acoustic transmission loss spectra with and without the guide valves, measured in the presence of 40m/s room temperature mean flow, are presented in Fig. 7. The aerodynamic pressure loss characteristics measured in 0-40m/s flow velocity range for both the silencer settings are shown in Fig. 8. The respective output torque curves measured in rolling road test facility for the motorcycle engine equipped with the silencer are exhibited in Fig. 9.

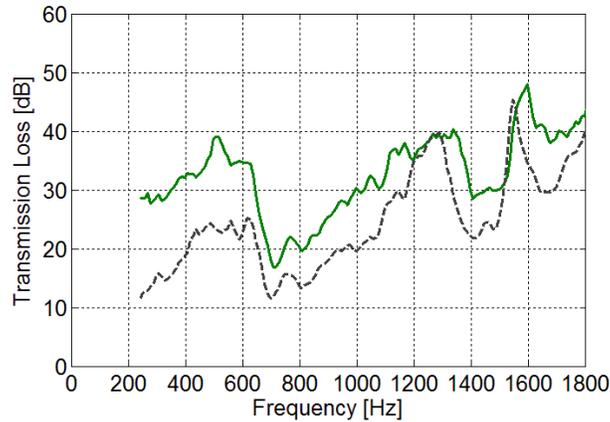

*Figure 6 – Transmission loss spectrums of the silencer measured in the presence of 40m/s mean flow velocity and presented for straight flow (grey dashed line) and high attenuation (green solid line) setup.*

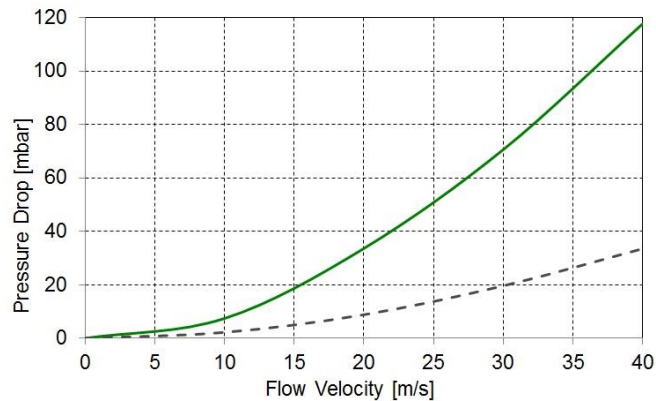

*Figure 7 – Experimentally determined pressure drop of the silencer presented for straight flow (grey dashed line) and high attenuation (green solid line) setup.*

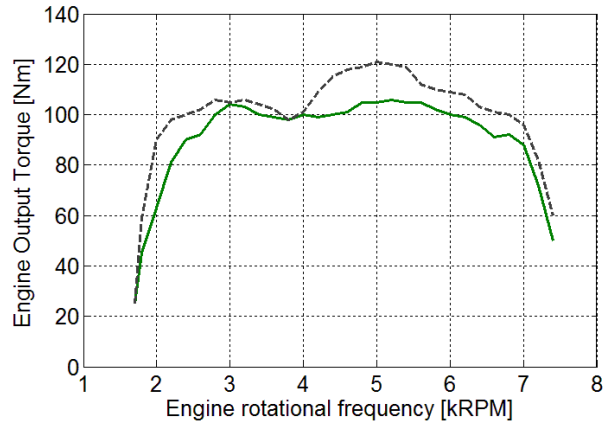

*Figure 8 – The output torque curves measured in rolling road test facility for the motorcycle engine equipped with the silencer in straight flow (grey dashed line) and high attenuation (green solid line) setup.*

# MODELLING RESULTS

In Fig. 9 the 1-D modeling results by using GT-Power™ and SIDLAB™ for the acoustic TL of the silencer in straight flow configuration (in the absence of the guide valves) are compared to the experimental ones in no-flow conditions.

It is clearly noticeable that the entire plane wave region shown is well represented by the 1D simulation. The result demonstrates that the use of the quasi 3-D elements for the chamber modeling allows to provide reliable results almost up to 1000 Hz.

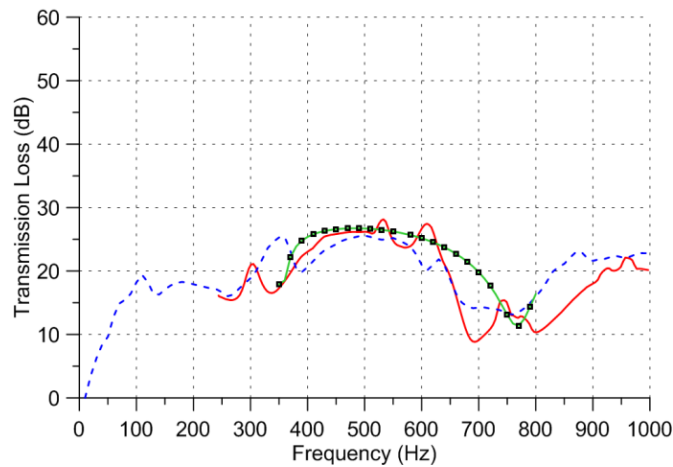

*Figure 9 – The acoustic TL of the silencer in straight flow configuration and in the absence of mean flow: experimental results (red solid line), 1D simulation results with GT-Power™ (blue dashed line) and SIDLAB (black dots).*

A comparison of the TL results for 40m/s mean flow velocity through the silencer in straight flow configuration (Fig. 10) and in case of the extra attenuation guide valves (Fig. 11) is presented. Due to the viscous effects generated by the flow passing the apertures, both the experimental and the numerical results obtained by GT-Power™ demonstrate an enhancement of the attenuation performance of the silencer at low frequencies. This effect is even more pronounced when the guide valves are equipped (see Fig. 11).

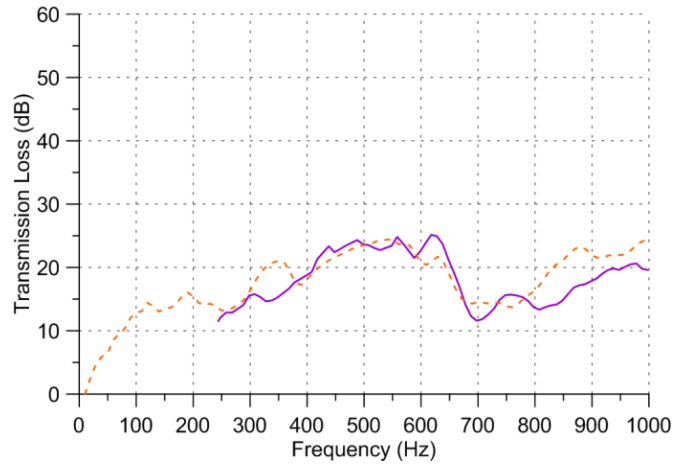

*Figure 10 – The acoustic TL of the silencer in straight flow configuration and in the 40m/s mean flow condition: experimental results (magenta solid line), 1D simulation results with GT-Power$^{TM}$ (orange dashed line).*

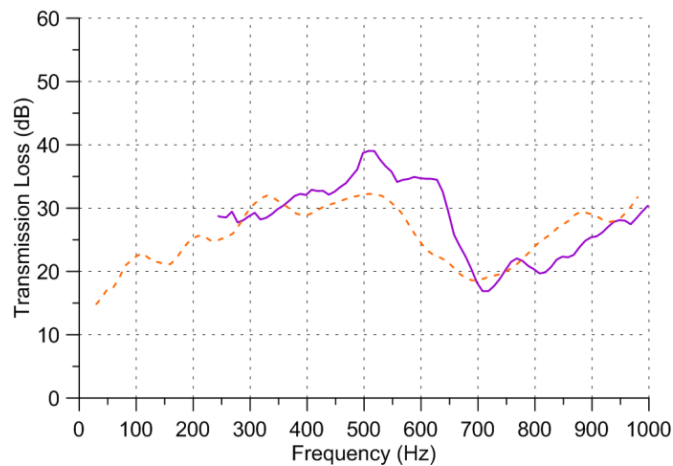

*Figure 11 – The acoustic TL of the silencer equipped with the guide valves and in the 40m/s mean flow condition: experimental results (magenta solid line), 1D simulation results with GT-Power$^{TM}$ (orange dashed line).*

The PD calculated by GT-Power$^{TM}$ simulations for two silencer configurations are 31mbar and 135 mbar. The simulated PD values agree well with the experimentally determined ones: 35mbar and 118mbar respectively (see Fig. 7). The relatively small difference between the simulated and measured PD of the silencer indicates well-captured viscous losses that occur in the presence of mean flow.

## CONCLUSIONS

A novel design for a cruiser type motorcycle silencer incorporating custom micro-perforated elements has been presented in this paper. The silencer represents an effective engine noise cancellation solution in constricted conditions without the implementation of the traditional fibrous materials.

The performance of the silencer as well as the engine output characteristics are presented for two different silencer configurations. The numerical 1-D simulation models developed for the optimization procedures have exhibited a good agreement with the experimentally determined data.

While maintaining acceptable pressure drop characteristics and providing pleasant engine sound, the silencer has proven to comply with standard noise criteria.

# CONTACT INFORMATION

Raimo Kabral: The Marcus Wallenberg Laboratory, The Royal Institute of Technology, Teknikringen 8, Stockholm, SE-10044, Sweden, kabral@kth.se, phone: +372 50 24 992;

Dr. Hans Rämmal: Department of Automotive Engineering, Tallinn University of Technology, Ehitajate tee 5, Tallinn, 19086, Estonia, hansra@kth.se, phone: +372 56 465 738;

# ACKNOWLEDGMENTS

The authors would like to acknowledge technical consultancy companies Lettore and Triple Seven for successful co-operation including financial, technical and technological support. In addition, the support from Estonian Science Foundation (Grant No. 7913), Marie Curie European doctoral student program and "DoRa" program of Archimedes Foundation are acknowledged.

# DEFINITIONS/ABBREVIATIONS

| | |
|---|---|
| **MP** | micro-perforated |
| **TL** | transmission loss |
| **PA Ratio** | perimeter to area ratio |
| **PD** | pressure drop |
| **RPM** | rotations per minute |